# Holographic Bragg gratings: measurements and examination diffraction parameters

E. A. Tikhonov, A.K. Lyamets

Institute of physics of NAN of Ukraine, Nauki Avenue 46, Kiev, 01680, etikh@iop.kiev.ua

**SUMMARY.** Measurement and analysis of diffraction parameters of thick transmission holographic phase gratings recorded on PHC-488 photopolymer are presented. Precision determination of the spatial grating period is executed by Bragg angle measurement for two precisely known wavelengths in the first diffraction order and also by Bragg angle measurement on one wavelength in the first and second diffraction orders. Determination of actual grating thickness and depths of spatial modulation of a holographic recording, which are responsible for their main optical characteristics, is carried out with the application of H.Kogelnik 2-wave coupled model. Calculated and experimental Bragg angles in two orthogonal orientated planes of linear polarized laser light were used to confirm theoretical model with experimental results. Extreme values of the diffraction parameters – efficiency and spectral resolution – important for holographic phase grating applications are analyzed. Work is performed due to the budgetary financing of Nation Academy Science of Ukraine, project number of state registration– 0106U002454.

## 1. INTRODUCTION

Applications of diffraction gratings in spectroscopy and laser physics are based on the use of the suitable optical characteristics derivative from the spatial period and the phase profile of a grating. Higher accuracy and efficiency of measurements by spectroscopic devices can be achieved with precise determination of grating parameters. In this work various aspects and procedures of precision measurement of periodic structure parameters belonging to the class of holographic phase gratings (HPG) and possessing perspective characteristics for applications in spectral devices for spectral area 350 - 1200 nm is considered /1/.

The quantitative description of HPG parameters might be effective in regards to the coupled wave model based on 2-wave interaction approach the incidence and the diffracted waves at harmonic spatial modulation of material index refraction on which a holographic grating is recorded /2/. The extent of the model to describe HPG parameters is sufficient, however, theory demands the corresponding experimental justification. Despite the advantages of the model, measurements of the spatial period of HPG is not considered within the model /2/ and are based on the unconditional application of the Bragg condition to light diffraction on the periodic structure.

For the description of HPG and applicability to them of the theory /2/ the following additional diffraction criteria are considered to be important: multi-wave and two-wave Cook-Klein's criteria /3,4/. Diffraction estimates in accordance with these criteria become possible at actual thicknesses of diffraction structures. Determination of the spatial period $\Lambda$ and thickness T in case of volume HPG cannot be executed directly. Opposed to surface gratings that can be studied with scanning microscopy /5/, volume and phase character of the HBG profile and the holographic record on photopolymer materials in real time accompanied with shrinkage, exclude the similar measurement. Therefore, determination of thickness relies on the measurement of some optical parameters - diffraction efficiency at a zero detuning from the exact Bragg angle under the condition of applicability to these HPGs Kogelnik's theory /2/.

Since the process of holographic recording is based on the two plane wave interference on the given wavelength λ, grating spatial period is known essentially already on a recording stage in accordance to classic ratio $\Lambda = \lambda/2n\sin(\theta)$. However the accuracy of similar determination $\Lambda$ at recording stage is limited due to small errors of installation of an external crossing angle θ, possible deviation bisector θ from a normal to grating orientation, inexact knowledge of wavelength and varying refraction index of holographic material used for record even in presence of precision measuring instruments of wavelengths and refractive index /6/. Especially in time recording of the reflective or transmission gratings with the inclined phase planes, when the period $\Lambda$ directly depends on the shrinkage and the refraction index of the registering material /7/.

Therefore, it is represented inexpedient to measure $\Lambda$ with high precision at the recording stage and to make similar measurement after manufacturing of gratings.

## 2. MEASUREMENT OF HBG SPATIAL PERIOD

Determination of the spatial period $\Lambda$ for manufactured Bragg type HBG can be executed with enough high precision using synchronized a goniometer of angular rotation (with a minute/second step) and system of photo-electron registration of angular dependence of the diffraction efficiency (DE) for the laser radiation of the well-known wavelength. Under Bragg conditions, the separately taken angular and wavelength errors summarize and their contributions can be considered approximately equal.

As an additional limitation to the high precision of period measurement by Bragg method the possible inclination of the phase planes to the grating surface acts. The inclination results in two external Bragg's angles of diffraction $\varphi_1$, $\varphi_2$ become different. It develops in the ambiguity of the spatial period. Ambiguity can be excluded by the following averaging of measured results. Let's summarize separated results of measurements which are reduced to receiving two different Bragg angles ($\varphi_1$, $\varphi_2$) and the two different periods ($\Lambda_1, \Lambda_2$). At the small tilt angle after elementary manipulation with classic ratio we receive a ratio between the averaging grating periods at an inclination of the phase planes and its efficient value:

$$\lambda = 2\Lambda_{ave}\sin\left[0,5\left(\varphi_1 + \varphi_2\right)\right]\cos\left[0,5\left(\varphi_1 - \varphi_2\right)\right] \cong 2\Lambda_{eff}\sin\left(0,5\left(\varphi_1 + \varphi_2\right)\right) \quad (1)$$

here angular half-sum defines Bragg's angle for not inclined phase planes and the half-difference of these angles is their tilt angle. The cos (…) member makes the amendment to the actual value of the spatial period $\Lambda_{eff}$.

The relation between the angular divergence of the laser beam and angular selectivity of DE studied HBG can also influence the accuracy of spatial period measurements. According to Kogelnik's results /2/ angular selectivity of the transmissive HBG is defined by the angle between the first left and right zero of DE described by sync-function that is approximately equal to $\Omega \cong 2L/T$. For grating thickness $T \approx (20 \div 100)\mu m$ and $\Lambda \approx 1\mu m$ angular selectivity to be within $(10 \div 50)$mrad while typical beam divergence of used gas,- or diode lasers with the correcting lenses is about or less 1 mrad. Respectively any noticeable broadening of an angular width of DE at given relation of divergences does not effect on the error of measurement. The detailed technique of maximum DE determination will be presented below.

Yet even, we proceed directly to measurements of DE angular dependences of the transmission HBG. According to the provided ratio (1) at a Bragg angle difference [0,5 ($\varphi_1$-$\varphi_2$)] ≤ 1' cosine member is differed slightly from 1 (just in 8th sign after a comma) so at such small phase plane inclinations it is possible to determine the grating period from a single measurement. The scheme of measurement includes a goniometer with the drive from the step motor, the controller and ADT/DAT drivers for determination of step and range of scanning and 2 photodiode receivers of radiation. At a small wedge in the grating sample constructed as triplex (glass-polymer-glass) and permanency of diffraction angles, photodiode receiver remains in invariable place under incidence angle variation of a rotation of samples. The photovoltaic mode of switching provides linear voltage response concerning light intensity variation within about 3 orders of values.

The suggested technique of $\Lambda$ determination is based on absolute measurements of Bragg angles in the presence of well-defined laser wavelength. It forces to set precision zero references of the goniometer, or others words - to set zero angular position of the diffraction grating plane. Though zero angular setting of the grating with a marginal error is commensurable that for Bragg angle measurements, reasonably to avoid the time-consuming operation. That is why we used following approaches for the solution of zero reference problem.

In the first one, two different well-defined wavelength lasers are obligatory. Arbitrary but the same initial angular position is chosen for both laser beams with the equal systematic error. Let's write down the corresponding system of equations for unknown $\Lambda$ and $\varphi_1$:

$$\lambda_1 = 2\Lambda\sin(\varphi_1) \qquad (2a)$$

$$\lambda_2 = 2\Lambda\sin(\varphi_1 \pm \varepsilon) \qquad (2b)$$

Having divided (2a) on (2b) we find expression for $\varphi_1$:

$$\varphi_1 = arc[ctg\,((1-(\lambda_1/\lambda_2)\cos(\varepsilon))\,/(\lambda_1/\lambda_2)\sin(\varepsilon))] \qquad (3)$$

Substitution (3) in (2a, 2b) allows to find the spatial period $\Lambda$ of given grating taking into account a difference of Bragg angles correspondent to maxima of angular dependence of DE for the chosen wavelengths. The need for evaluation system error of zero references in the case is eliminated. As for the method of precise determination of DE angular maximum, we will refer to the similar technique that contains in computer software like "Origin" and some similar.

Another possibility of precise determination of the grating period $\Lambda$ of considered HBG appears if the second diffraction order is observed. This question is closely connected with applicability degree of the theory /2/ to our photopolymer HBG. Meanwhile, determination of the spatial period by Bragg's method, existence and the nature of the higher diffraction orders has no evident effect. Here it should be noted also that many times lower

DE in the higher diffraction orders can be registered if the increased Bragg's angles do not result in TIR of diffracted waves inside grating volume. Besides, DE in the higher diffraction orders falls down with the growth of grating thickness within $(10\div100)\mu m$ where Klein-Cook's condition /3/ defines the border of the existence of 2nd and 3d multi-wave diffraction. Observation of the second (and higher) diffraction orders is defined just by Bragg's condition, and DE in the corresponding orders depends on an anharmonic profile of holographic record and probability of two-photon scattering in general.

Let's consider conditions of Bragg's diffraction in higher orders. They are subjected to the law of momentum conservation at elastic light scattering of the coming and diffracted radiation $k_{in}$, $k_s$ on a grating with $K=2\pi/\Lambda$ wave-vector. At multi-wave diffraction scattering angles concerning the direction of the incident wave is small because $K<<k_{in,s}$. While at the two wave diffraction scattering angles are bigger because $K\approx k_{in,s}$. In a vector form of representation, the momentum conservation law for multi-wave and 2-wave diffractions do not differ. The vector form $k_s \pm k_{in} = mK$ transforms into the known equation of diffraction grating:

$$\Lambda(\sin(\alpha)\pm\sin(\varphi))=m\lambda, \qquad\qquad (4)$$

where m=1,2,3 … an order of diffraction, $\alpha$, $\varphi$- incidence and diffraction angles respectively. The grating equation at big and equal incidence angles and diffraction takes the form of a Bragg condition.

Thus, multi-wave diffraction is possible owing to the bigger probability of small changes of an impulse of the incident light at addition with small grating impulses K (m value can change from 1 to ~ 20). At Bragg diffraction, when scattering impulses of light and a grating are nearly equal, therefore scattering on big angles happens with high probability only for first diffraction order m=1; at diffraction in 2-3 orders scattering angles increase so strongly that resonant conditions are not always feasible. The quantitative ratio for conditions of alternative existence multi-wave and 2-wave Bragg diffraction on the ultrasonic wave gratings was received in early work /4/, then later in Cook-Klein' work /3/:

$$Q=2\pi\lambda T/n\Lambda^2 \qquad\qquad (5)$$

Here n is the refractive index of the material with spatial modulation period $\Lambda$ and T - grating thickness. Experiment and the theory show in agreement that at values $Q\leq0,3$ multi-wave diffraction is observed, at $Q\geq4\pi$ two wave Bragg diffraction is developed. In a consent with (5) in the intermediate range of values, Q up to 1 order of values is allowed the simultaneous existence of two-wave and multi-wave diffractions.

Taking into account the purpose of this work – precise determination of parameters $\Lambda$ and T - we will consider area of existence of the allowed angles of Bragg's diffraction and, respectively a possibility of determination of the spatial period by way of direct Bragg's angle measurement in first and second diffraction orders without the need of zero count setting of a goniometer. Substituting m=1 and 2 in the equation (4) (Bragg's condition at equal incidence and diffraction angles) we will find an angular difference for 1 and 2 diffraction orders:

$$\varphi_2 - \varphi_1 = \varepsilon = \arcsin(\lambda/\Lambda) - \arcsin(\lambda/2\Lambda) \qquad /6/$$

The solution of the equation (6) for given values of $\lambda$ and the measured value $\varepsilon$ allows to determine the space period with an accuracy limited only by measurement errors of Bragg angles and wavelength of the used laser. For an illustration of the method in fig. 1. calculated dependence of diffraction angles in the first five orders is provided.

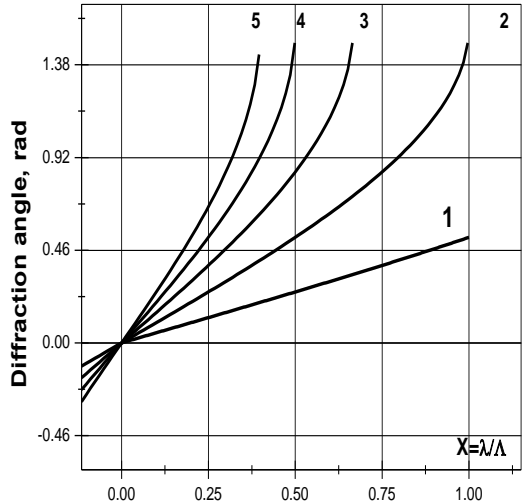

Fig.1. Dependence of diffraction angles on $\lambda/L$ relation in the higher orders of Bragg diffraction (the 1-st, 2,3,4,5 - orders).

From fig.1. it follows that each $\lambda/L$ value corresponds the definite value of Bragg angles in the orders first, second, etc. For values of relation $\lambda/L < 0.6$ multi-wave diffraction is allowed, while for $\lambda/L > 0.6$ number of orders decreases to no more than 2. For $\lambda/L \equiv 0.9$ values and more Bragg diffraction in the second order for a grating with inclined phase planes gets unobserved because beam localization inside the grating layer. The equation (6) has been solved digitally for the unknown grating period $\Lambda$ at the measured value $\varepsilon$ and for the given wavelength $\lambda$. Let us show $\Lambda$ measurement on an example He-Ne and He-Cd lasers. Fig. 2. represents DE angular dependence in the 1-st and 2-nd diffraction orders for the symmetric Bragg phase transmission grating.

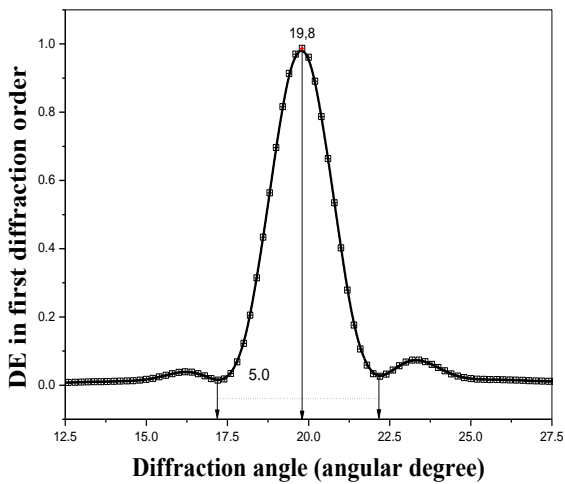

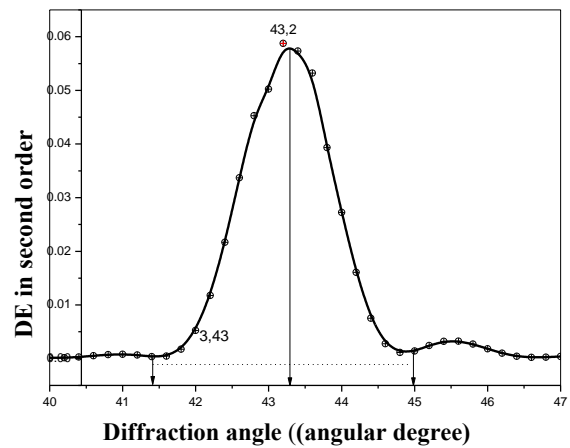

Fig. 2. Angular dependences DE HBG for HeNe laser ($\lambda$=632,8nm, s-polarization, sample No. 4) in first (left) and second (right) diffraction orders. The relation $DE_{max}(1)/DE_{max}(2) \cong 17$ and increases with grating thickness T.

The angular position of both $DE_{max}$ are determined with a margin error 1' (depends on goniometer type). This measurement does not contain an error of zero count setting of our goniometer. Presented on fig. 2. angular values of $DE_{max}$ reflect results of their standard determination by Origin 8.5 soft. There is an opportunity to increase the determination accuracy of $DE_{max}$ to fractions of an angular minute, but we will not concern this detail. In the case under consideration the difference of $DE_{max}$ angular positions in first and second orders for the given grating makes $\varepsilon$=43,29-19,80=23,49° (0,1305rad). The numerical solution of the equation (6) for this $\varepsilon$ and $\lambda$=632,8nm allows to find the grating spatial period No.4 $\Lambda$=915,96nm. The number of signs after a comma is limited to the wavelength accuracy which is typical for He-Ne laser without frequency stabilization. At the usage lasers with the stabilized frequency the measurement accuracy $\Lambda$ can be respectively increased. In our work /7/ it was offered to use similar transmission HBGs with calibrated spatial period for precise measurement a wavelength of light beam sources.

## 3. ABOUT VALIDITY OF KOGELNIK'S MODEL TO PHOTOPOLYMER HBG

The next step of the work consists in the determination of post-shrinkable thickness T and modulation depth $n_1$ of photopolymer gratings at real time recording. As it was noted above, a basis of such assessment serves experimental measurements of HBG parameters and their processing with the application of theoretical model /2/. Grating thickness T and modulation depth $n_1$ are connected in the theory through measured $DE_{max}$ values and DE angular width defined by 2 first zero values of sync- function (fig. 2.). The theory is built in the assumption of the harmonic and polarizing isotropic response of record on the used photopolymer material. However, an observed anharmonic response that is proved by available diffracted radiations in the 2-nd order of diffraction already causes concern about the rather unconditional feasibility of the first of the called approximations. Measurement of radiation in the second order diffraction with different thickness gratings shows that its power (at equal $DE_{max}$) decreases in proportion to thickness T. It means that $n_1$ is inversely proportional to modulation depth and similar behavior of a response to perturbation is typical one for the interaction of radiation with substance. If on grating sample No.4 at T=15 $\mu$m the ratio of diffraction power in first and second orders is equal 17, while, at a thickness of 100$\mu$m and more, standard registration of diffraction power in the second order becomes difficult due to an essential decrease of power. Therefore, the approach of a harmonic response for the analysis of parameters of thick photopolymer HBG recorded on PPC-488 type can be considered satisfactory.

The second approach of theory /2/ stands on isotropy of a response can be broken and developed in the distinction of amplitudes of spatial modulation of $n_1$ for two orthogonal oriented linearly polarized radiation (s, p-waves). Let's apply to the analysis of the feasibility of this approach. DE measurement of HBG for s-, p-orientated linearly polarized radiation for which the theory /2/ predicts different $DE_{max}$ values at equality of amplitude modulation of $n_1$ forms the basis of similar analysis. Distinction $DE_{max}$ for these polarizations in the theory /2/ is connected with distinction of values of phase shifts for $v_{s,p}$:

$$DE(s)= \sin^2(v_s) , \quad v_s =\pi n_1\lambda\cos(\theta_{in}) \qquad (7a)$$

$$DE(p)=\sin^2(v_p) , \quad v_p= n_s \cos(2\theta_{in}) \qquad (7b)$$

here $\Theta_{in}$ - an internal Bragg angle which cannot exceed TIR angle equals $\approx 41^o$ for the refraction index n=1,53. As phase shift for p-polarization lags behind its value for s-polarization, the distinction of DE(p) concerning DE(s) increases proportionally the value of this shift and becomes maximum at $v_s=\pi/2$. It is practically possible to record gratings with Bragg incidence angles $57^o$ which is equal Brewster angle ($\Theta_{brs}$) at value n=1,53. Respectively even at grating recording at a Brewster angle the internal Bragg angle will not exceed value $33^o$.

All subsequent measurements of the grating DE are performed as much as possible to decrease losses of Fresnel reflection and small absorption/scattering in HBG layer. Diffraction efficiency was defined as DE = $P_{dif}$ / ($P_{dif}$ + $P_{trans}$), where $P_i$ - powers of the diffracted and transmitted beams of laser radiation. Typically for all available HBG samples at a phase shift of $v_s \leq\pi/2$ DE(s)> DE(p), but here is required a quantitative evaluation of their relation. Let's define internal Bragg angle from system equations (7):

$$\cos(2\theta_{in})=\arcsin(DE(p))^{o,5}/\arcsin(DE(s))^{o,5}=(n^2- \sin^2(2\theta_{out}))^{o,5}/n \qquad (8)$$

The external Bragg angle $\Theta_{out}$ is measured with a high accuracy and can be found from expression (8).

$$\theta_{out}= o,5\arcsin\{n[1-(\arcsin(DE(p))^{o,5}/\arcsin(DE(s))^{o,5})^2]^{o,5}\} \leq\theta_{brs}=57^o \qquad (9)$$

Thus, (9) allows to find external Bragg angle through the measured DE values. The coincidence of the calculated and measured external Bragg angles will be the proof of polarizing isotropy and, respectively, the proof of applicability of the theory /2/ to our photopolymer holographic gratings. Unfortunately, ln this approach are a necessary measurement of the third unknown − index of refraction for the recording material after HBG manufacturing.

By analogy with the definition of an absorption dichroism lets determine degree of holographic recording isotropy like next:

$$I_{nso} \sim (\theta_{outm} - \theta_{outt})/( \theta_{outm}+ \theta_{outt}) \qquad (10)$$

where external calculated by (9) and measured Bragg angles are used.

Measurements of the refraction index of the polymer layer for HBG recorded on PPC-488 by Brewster refractometer /8/on wavelength 632,8nm Ne-Ne laser has shown n=1,56. In fig. 3. DE (s,p) of grating with a spatial frequency (908,55 mm$^{-1}$), are provided in vicinities of a maximum of DE for the angle $20,388^o$. The calculated value of Bragg angle for DE values on two polarization (DE(s)=0,815, DE(p)=0,710) equals $20,110^o$ that was received at the numerical solution of the equation (9). The small anisotropy degree at the same time is equal to 0,69%. Calculation procedure naturally indicates the strong dependence of a calculated Bragg angle from the DE measurement accuracy  and magnitude of index refraction n.

Larger  distinction in DE magnitudes for s-, p-polarization with increase  Bragg angle is expected for gratings with higher spatial frequency. Possible recording anisotropy for a such gratings occurs to be stronger. In fig. 4. the measured data on DE(s)=0,97 and DE(p)=0,71 for the grating with a spatial frequency 2255mm$^{-1}$ and external Bragg's angle 45,44$^o$ on 632,8nm respectively are provided. It is necessary to make the amendment to a value of  DE(s) at these incidence angles because of increase of reflection on an output plane of a substrate that provides to the growth of DE(s) $\cong$1 at the accepted processing of observed data. Moreover, there is some increase of refraction index of  the registering polymer layer owing to more than 2-fold increase in spatial frequency (in the given example - n$\approx$1,57). Substitution of the measured values in a formula (9) results in $\Theta_{outt}$=44,11$^o$ value. The degree of anisotropy of recording at the same time increases to 1,48%.

It is necessary to notice that the higher accuracy to measurements of DE and refraction index  (availability of 3 signs after a comma)  lacks for exact calculations and conclusions with formula (9). Nevertheless at a more low spatial frequency, the concurrence of calculated and  measured Bragg angles is enough  for a conclusion about the feasibility of approach of  isotropic holographic recording response on the photopolymer  PHC-488. Correspondingly, the conclusion about calculation eligibility  of polymer HBG parameters on the basis of Kogelnik's  model is justified/2/.

## 4. DETERMINATION OF THICKNESS AND  MODULATION DEPTH OF HBG

The above-presented results give  grounds for a determination of the actual thickness of T and modulation depth $n_1$ for transmission HBG recorded on PHC-488 photopolymers. The theoretical model /2/contains necessary solutions for a numerical evaluation of the main HBG optical parameters: angular, spectral and polarizing variations  of DE, absolute values DE$_{max}$, etc. At applicability of the theory to  numerical evaluation of the HBG "hidden" parameters can be calculated already indirectly as a solution of the return task. The knowledge of the main "hidden" $n_1$ and T parameters is necessary for an understanding of the recording, opportunities of application of HBG of this class and optimization of their parameters.

DE of  transmission HBG for s,- and p-type polarized waves /2/can be presented in the following convenient for the analysis form:

$$\eta(s)=\sin^2(\nu_s\{1+[(n/n_1)\delta\sin(2\theta)]^2\}^{0,5})/1+[(n/n_1)\delta\sin(2\theta)] \qquad (11a)$$

$$\eta(p)=\sin^2(\nu_P\{1+[(n/n_1)\delta\tan(2\theta)]^2\}^{0,5})/1+[(n/n_1)\delta\tan(2\theta)]^2 \qquad (11b)$$

Let's determine the magnitudes $n_1$ and T HBG, relying on analytical sync- function properties of above-provided expressions: here 2$\delta$ - the angular width of DE$\equiv\eta$, at $\delta$=0 $\eta$(s), $\eta$(p) reach maxima at the set internal Bragg angle  (7a, b). Though the equations (7a, b) do not form a total system of unknown $n_1$ and T, they allow to find their production ($n_1$T) concerning measured values DE and Bragg angle set by the period Λ.

The missing link for formation and calculations total system of equations about $n_1$ and T for can be created when to equate zero (11) that is true by detuning from an exact Bragg angle on $\delta_{\pm 1}$:

$$\delta^s_{\pm 1} = [(n_1/n)[(\pi/v_s)^2 - 1)]^{0,5}/\sin(2\theta) \qquad (12a)$$

$$\delta^p_{\pm 1} = [(n_1/n)[(\pi/v_p)^2 - 1)]^{0,5}/\tan(2\theta) \qquad (12b)$$

$2\delta = \delta_{+1} + \delta_{-1}$ is the angular width of DE between the first symmetrical zeros of the DE-sync-function.

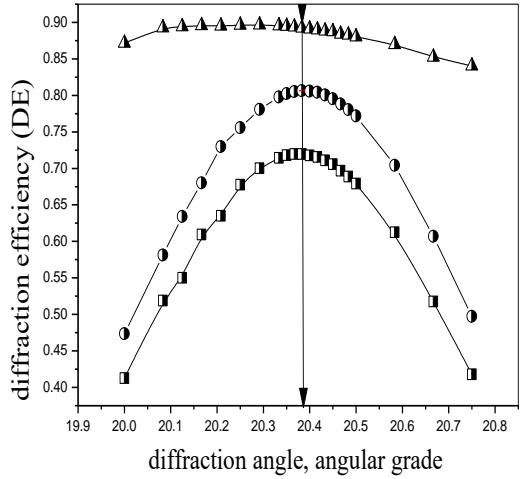 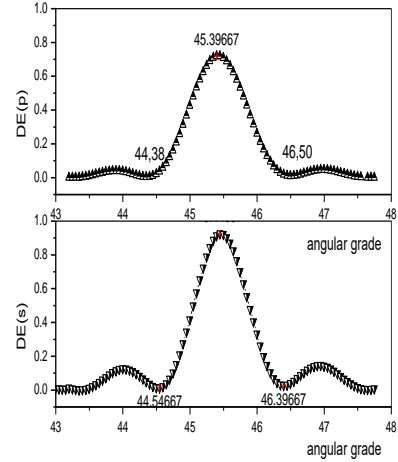

Fig. 3. Grating DE(s) –top and DE(p)-dawn for s,- and p-polarization and their relation (most upper)- as a function of the diffraction angle. An exact Bragg angle = 20,388º at λ=632,8nm.

Fig. 4. Grating DE with a spatial frequency 2251,3mm⁻¹ for p (top) and s (bottom) polarization, DE(s) =0,922, DE (p) =0,729.

At further processing of the equations (12) and transition to external Bragg angle, which is measured directly, substitution of the $v_s = \pi/2$ value and $v_p = (\pi/2)\cos(\Theta)$ corresponding $DE_{max} = 1$, we will receive two ratios for calculation of modulation depth of refraction index $n_1$:

$$\delta_{\pm 1} = 3n_1/\sin(2\theta_{out}) \qquad (13a)$$

$$\delta_{p\pm 1} = [3n_1/\sin(2\theta_{out})](1 + \sin^2(2\theta_{out})/3n^2) \qquad (13b)$$

The defined result specifies that width of angular selectivity for s, - p-of polarization at DE(s)=1 differ and this difference is in percent shares.

However, we are interested in an evaluation of thickness T and $n_1$ at an arbitrary value of DE. With the formulas (12a, b) transformed to external Bragg angle, we come to determination of geometrical thickness for s, p- orientations of field polarization:

$$T=\Lambda/\delta_{s\pm1}[n^2+n^2_1/\ \delta_s^2\pm1\sin^2(2\theta_{out})]^{0,5} \tag{14a}$$

$$T=\Lambda/\delta_{p\pm1}[n^2+n^2_1/\ \delta_p^2\pm1\tan^2(2\theta_{out})]^{0,5} \tag{14б}$$

It is obvious that the values T defined from the expression (14) should not differ for material with an isotropic response. Comparison of thickness on formulas (14) can serve as another test of applicability of the theory /2/ to the used holographic material. A difference in angular selectivity of DE s,- and p-polarizations for our transmission grating fig. 4. makes $(6\div7)$'. The equations (14) allow to determine the geometrical thickness T with the degree of accuracy depending on measurement errors of $\Lambda$ and $\delta_{\pm1}$. The member in square brackets (14) provides the grating optical thickness depending on refraction index n and the small correction. The procedure of Brewster type measurement of n for the hologram grating is described in /8/.

The registration DE curves with well-defined zeros in points $\delta_{s,p\ \pm1}$ which angular positions is measured by the goniometer, allows to find out $n_1$ from the system (14a, b). After simple transformations (14) we come to convenient for evaluation expression:

$$n_1 = (\delta_p^2 - \delta^2_s)^{0,5} \tag{15}$$

On the case HBG with data in fig. 4. : $\delta_p = 0,0194$rad, $\delta_s=0,0174$rad one finds out $n_1=0,0084$. For values of the spatial period $\Lambda=0,444$mkm and index of refraction of $n=1,57$ the geometrical thickness of the specified grating according to (14) takes the value: $T_0=\Lambda/\delta_{s,p} = 31.37$ µm. We can receive value for an optical thickness taking into account the member in square brackets that gives T =49,25 µm. Let's check the obtained data by calculation of DE(s) with formula (12a) and makes comparisons to experiment: $\eta^s_{max}=\sin^2(2\pi*0,0083*31,37*\sin(45,40)/0,6328*\sin(90,8)=0,926$ that differ from directly measured DE(s)=0,922 value for the grating only in the third sign (fig. 4.). Thus, calculation of recording modulation depth $n_1$ and thickness of a diffraction layer T can be carried out by correct measurement of DE even on the single type of polarization.

A bit different way of finding of the values $n_1$ and T at measurement of angular position of DE maximum. We will define the calculated value of $DE_{max}$, differentiating the equations (7a, b) on $\nu_s$ and equating result to zero. Respectively we receive:

$$(n_1T)_{max}= \lambda\sin(2\theta_{out})/2\sin(\theta_{out})=\Lambda\sin(2\theta_{out}) \tag{16a}$$

If the measured size DE differs from 1, then product $(n_1T)$ in DE maximum is found out:

$$(n_1T)_{эmax} = [\lambda\sin(2\theta_{out})/2\pi\sin(\theta_{out})]\ arcsin(\eta^s_{max}) \tag{16b}$$

or with the equivalent equation for DE(p).

Substitution of formula (15) in (16) give an expression for calculation of T grating by means of three angular parameters: $\delta_p$, $\delta_s$ and $\Theta_{out}$.

$$T= \lambda\sin(2\theta_{out})/2\sin(\theta_{out})(\delta_p^2 - \delta^2_s)^{0,5} \tag{17a}$$

$$T=[\lambda\sin(2\theta_{out})/2\pi\sin(\theta_{out})]arcsin(\eta^s_{max})/\ (\delta_p^2 - \delta^2_s)^{0,5} \tag{17b}$$

Product $(n_1 T)_{max}$ defined by (16) corresponds to DE maximum in the first diffraction order. Apparently from (16a), its extreme value does not exceed grating spatial period $\Lambda$ or record wavelength $\Lambda/2,82$ for Bragg's angle $45^o$. It allows to estimate extremely achievable values of the thickness of similar HBG which are possible in the experiment and, respectively, the extreme value of angular selectivity of the transmission HBG /10/. When at grating recording the optical thickness exceeds specified, DE on given wavelength passes maximum, then falls down to zero (in fact to some small values), then again increases, but never reach 100% in an experiment against to the calculation in the frame of the model /2/. There are many reasons for rejection from the theoretical scenario of DE-evolution in a time of recording, but their analysis is outside this work.

## 5. ABOUT MARGINAL VALUES OF HBG PARAMETERS $n_1$ AND T

Our definite interest consists in what minimal values of modulation depth $n_1$ of refraction index n can be realized under conservation of DE maximum in the first diffraction order and conservation of above-received conclusion $(n_1 T) \approx 0,18 \mu m$. For typical $n_1$ values 0,001 most characteristic thickness of gratings, T is (0,1-0,2) mm.

The photopolymer PPC-488 allows to reduce the depth of index modulation $n_1$ of registering material n at record in several ways: with reduction of the concentration of a diffusing additive (the substance determining the modulation depth and long time stability of holographic recording) and initial partial polymerization of the main photopolymer just before recording /9,10/.

The lowest level of $n_1$ at holographic recording can be limited by thermal fluctuations of the refraction index of the holographic material. Though the metrology of refraction index allows to register their values within a margin error $10^{-5} \div 10^{-6}$ (when using goniometric methods and $10^{-7} \div 10^{-8}$ - with interferometric methods /11/), these results belong to averaged values.

It is also known that the greatest contribution to the power of the scattered light is made due to static inhomogeneity of refraction index of materials. The theory of this scattering has been developed by Mi for the scattering centers of all sizes $x \leqslant 1$ /11/. Static heterogeneity is connected with technology (or natural origin) of optical material manufacturing while fluctuations of the dynamic nature belong to nonremovable physical material properties: the spectral triplet of Rayleigh scattering results in fluctuations of entropy, density, and anisotropy. This contribution to integrated light scattering is significantly less.

Light scattering on quasistatic inhomogeneity of refraction of the holographic polymer is manifested brightly at hologram record of in real time: noise holograms follow the image recording especially brightly in real time record in liquid photopolymer phase /12/. Thus, in the analysis of similar task, it is necessary to take into account along with main plane waves that are responsible for main grating recording, likewise scattering waves taking place into the recording of casual noise holograms. If the record of casual holograms does not happen, the small amplitude of regular record $n_1$ will be subject to indignations of above-mentioned fluctuations of refraction index. At the same time, when the spatial harmonic of fluctuations will become comparable on the spatial frequency of the main grating even in linear approach their addition can decrease DE of the hologram grating.

Therefore, the lower limit of the value $n_1$ will be limited  by fluctuations of refractive index n of holographic material at the set temperature. Total amplitude of fluctuations of air refraction index  in the typical conditions is equal $\approx 3*10^{-4}$ /13/. The level of light scattering R=$P_{sct}$/$P_0$ can change from 100% for materials with multiple scattering to (0,01÷0,001) % for quartz glasses used in light fibers. At record-breaking low losses 1db/km for optical fiber, scattering  coefficient  is equal $\alpha\approx 0,1$ km$^{-1}$; for polymer optical fibers $\alpha$ is 1-2 orders anymore /14/. Assuming that light scattering on refraction index fluctuations n is a major factor of losses on a light transmission for polymers, it is possible to estimate the maximum permissible thickness of grating T can be evaluated as $0,18\mu m/10^{-(3...4)} = (0,18÷1,8)$ mm. Extreme thickness of holographic grating T on  our photopolymer  PPC-488 at saving of grating DE$\approx$100%  reached 1 mm /10/.

## 5. SPECTRAL DEPENDENCE OF DIFFRACTION EFFICIENCY  HBG

The most important feature of HBG is the spectral dependence of inherent DE the in working frequency range. For volume phase gratings,  the spectral transparency range nominally matches a band of transparency of the used photopolymer. The  spectral transparency window for PPC-488 photopolymer overlaps  interval (350÷1200)nm. Formulas (7) specify that 100% DE are reached at a phase shift of $\pi$/2, however, this optimal value does not remain constant within the working range for two reasons: owing to frequency dependent phase shift of $\nu_{s,p}$ and owing to the normal dispersion of refraction index x=f($\lambda$). It is possible to accept as a first approximation that dispersion of modulation depth $n_1$=f($\lambda$) of acrylic polymer not significantly differs from the dispersion of bulk refraction index  of main polymer material. The known empirical formula dispersion in a transparency band  many organic materials is well described by Cauchy's relation: $n(\lambda) = (A + B/\lambda^2 + C/\lambda^4)$, where A, B, C - pilot parameters /6/. Coincidence accuracy to $10^{-4}$ is provided when using only of the first two members of a row. For a quantitative analysis of behavior conditions of  DE$_{max}$ (7) we will write down them in an explicit form excluding Bragg's angle:

$$\text{DE}_{max}(s)= \sin^2(\pi n_1(\lambda)T/\lambda\cos(\ \theta_{in})=\sin^2(\pi n_1(\lambda)T/\lambda(1-\lambda^2/4\Lambda^2)^{0,5}) \qquad (17a)$$

$$\text{DE}_{max}(p)=\sin^2(\nu_p),\ \ \nu_p=\nu_s(1-\lambda^2/\Lambda^2)^{0,5} \qquad (17б)$$

In fig.5. are shown a calculation of changes DE$_{max}$(s)  without taking into account dispersion of $n_1$=f($\lambda$) in a PPC-488  spectral transparency window. The curve 1 indicates that if to record HBG with DE$_{max}$ on the short-wave part of a spectral window - 0,3$\mu$m, then at the long-wave edge (1,2mkm) DE will fall 1 order of max value. The curve 5 specifies that HBG with a maximum DE approximately in the middle of the transparency window 0,65$\mu$m,  then at long-wave edge DE$_{max}$ discovers an only double decrease in efficiency while on wavelength 0,34$\mu$m DE(s) in general rushes to zero. The best choice of  phase shift value  ($n_1$T)= 0,3 occurs to be for dependence 3 with $DE_{max}$ on 0,5mkm: at edges of a transmission band DE does not fall down lower than 30%.

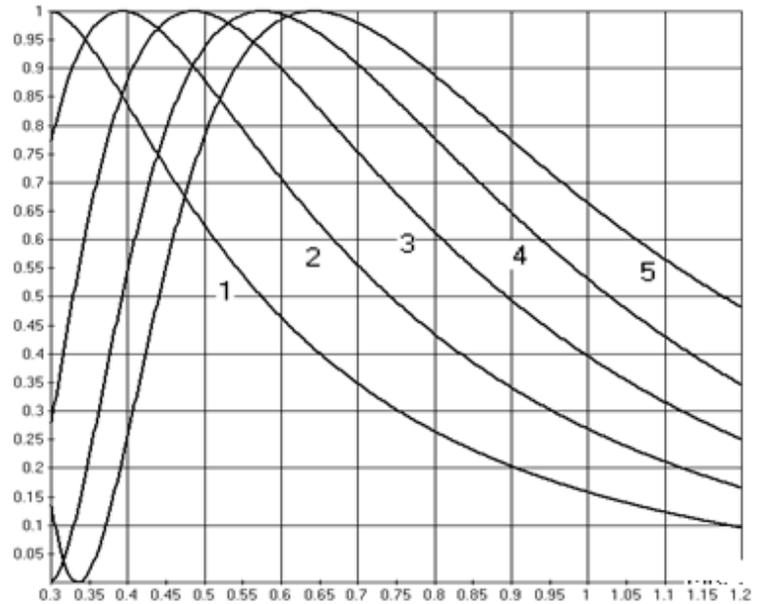

Fig. 5. Variations of $DE_{max}(s)= F(\lambda)$ at different values $(n_1T)$: 1-0,15; 2-0,20; 3-0,25; 4-0,30; 5-0,35; without taking into account dispersions of $n_1=F(\lambda)$.

High-quality coincidence to the described variations of DE shows an experiment with the grating which $DE_{max}\approx100\%$ intentionally was made at wavelength 0,632μm.

The same time the main maximum of DE on wavelength 0,441μm becomes significantly less on both sides DE curve due to so-called "overmodulation" when phase shift on given wavelength gets more $\pi/2$ (fig. 6). The fig. 5. specifies that to this case best solution corresponds to the curve 5 with DE(s)=0,4 against 0,12 on experience.

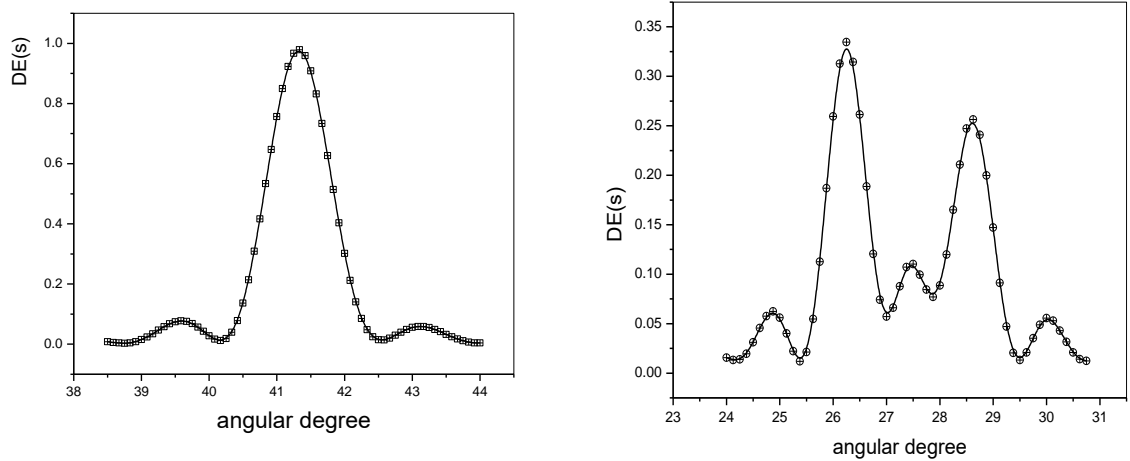

Fig. 6. Angular dependences of DE(s) HBG on wavelengths 0,632μm (at the left) and 0,441μm

However calculation of variations $DE_{max}$ in fig. 5. does not take into account a normal dispersion of refraction index. For an accounting of the last we will use 2-member Cauchi approximation in the form of $n_1T=(a+b/\lambda^2)T$. With reference to already made calculation of variations of $DE = f(\lambda)$ at $n_1(\lambda)T$ =const (fig. 5.) let's consider changes of $n_1(\lambda)$ in the same spectral range for PPC-488 photopolymer. The refraction index n of methylnaphthalene at 20°C at edges of a range (0,436 - 0,667)μm has values respectively 1,656 - 1,608, i.e. total change makes magnitude $\Delta n$=0,048. Because the diffusing additive of PPC-488 responsible for holographic recording on photopolymer is bromonaphthalene, dispersion of both naphthalene derivatives is proposed to be equal. Applicable changes of $\Delta n_1$ in the same spectral range can be considered proportional to the value of $\Delta n$. In that case 2-member Cauchi approximation for dispersion of modulation depth takes a form:

$$n_1T=(A+B/\lambda^2)T=0,15 + 10^{-3}/\lambda^2 \qquad (18)$$

Substitution (18) in a formula (17a) with a graphical presentation of the relevant calculation of DE(s) are shown in fig. 6. It can see that corrections to phase shift due to $n_1(\lambda)T$ dispersion at their values $10^{-3}$ do not exert impacts (dependences 1,2). A hypothetical increase of dispersion contribution up to $10^{-2}/\lambda^2$ manifests in narrowing of bands $DF(\lambda)$ – for curves 3,4 especially strong from the short-wave side. Dependence 5 shows change $DE(\lambda)$ at the increase in grating spatial frequency from 1000mm⁻¹ to 2000mm⁻¹. The narrowing of $DE(\lambda)$ function on the case is caused by known dispersion increase due to the diffraction structure.

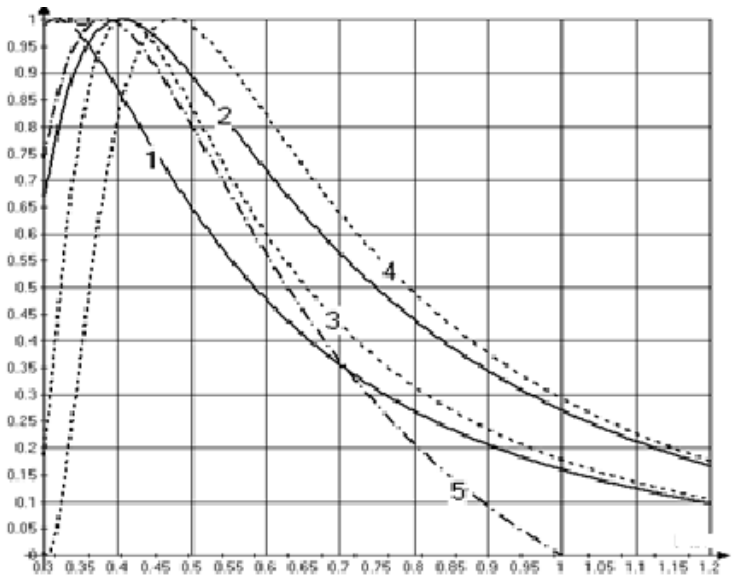

Fig. 7. The changes of $DE_{max}(s)=f(\lambda)$ at taking into account dispersion of modulation depth $n_1T$: 1-$(n_1T)$ =0,15+10⁻³/$\lambda^2$; 2- and 5- 0,20+10⁻³/$\lambda^2$; 3-0,15+10⁻²/$\lambda^2$; 4- 0,20+10⁻²/$\lambda^2$; for cases 1,2,3,4 spatial period of $\Lambda$=1μm; 5- $\Lambda$=0,5μm.

## 6. CONCLUSION

In given work optical parameters of the holographic Bragg gratings, recorded in real time on the phase holographic photopolymer, are measured and analyzed. Their reasonable analysis is provided on the basis of the theoretical model of the coupled waves offered by X.Kogelnik's work. The following statements of work represent its main results:

• Two methods of precisional calibration of spatial frequency of HBG excluding installation of a zero point of Bragg angle measurement are offered and implemented: a) by measurement of the first Bragg angles on two knowns (with a required accuracy) wavelengths. and b) by the measurement of the first and second Bragg angles at the single known wavelength

• The checking algorithm of quantitative ratios between the measured and hidden parameters of a holographic recording is proposed. The modulation depth of refraction index and an effective optical thickness of a grating are considered to the grating parameters hidden from direct measurement. It has been gained the positive conclusions about the applicability of the coupled waves approach theory in model X.Kogelnik for the analysis of the hidden optical parameters of the transmission type HBG on photopolymer composite PPC-488.

• Qualitative analysis of the physical factors defining extremal values of modulation depth of refraction index at HBG recording is performed. The smallest modulation depth limits greatest angular (and spectral) selectivity of HBG. Spatial fluctuations of the bulk refraction index of photopolymer are presumably referred to a number of such factors.

• The spectral position of maximum DE and its bandwidth of HBG in a spectral window of the maternal photopolymer is caused by normal dispersion of the registering material and dispersion of HBG phase shift. It is shown that extent of influence the phase shift dispersion on DE is prevailing.